\documentclass[prl,onecolumn,amsmath,amssymb,superscriptaddress]{revtex4-1}
\usepackage{bm,psfrag,graphicx,setspace}
\usepackage[left]{lineno}
\selectfont

\begin{document}
\title{Direct Observation of Infrared Plasmonic Fano Antiresonances by a Nanoscale Electron Probe}
\date{\today}
\author{Kevin C. Smith}
\affiliation{Department of Physics, University of Washington, Seattle, Washington 98195, USA}
\author{Agust Olafsson}
\affiliation{Department of Chemistry and Biochemistry, University of Notre Dame, Notre Dame, Indiana 46556, USA}
\author{Xuan Hu}
\affiliation{Department of Chemistry and Biochemistry, University of Notre Dame, Notre Dame, Indiana 46556, USA}
\author{Steven C. Quillin}
\affiliation{Department of Chemistry, University of Washington, Seattle, Washington 98195, USA}
\author{Juan Carlos Idrobo}
\affiliation{Center for Nanophase Materials Sciences, Oak Ridge National Laboratory, Oak Ridge, Tennessee 37831, USA}
\author{Robyn Collette}
\affiliation{Department of Materials Science and Engineering, University of Tennessee, Knoxville, Tennessee 37996, USA}
\author{Philip D. Rack}
\affiliation{Department of Materials Science and Engineering, University of Tennessee, Knoxville, Tennessee 37996, USA}
\affiliation{Center for Nanophase Materials Sciences, Oak Ridge National Laboratory, Oak Ridge, Tennessee 37831, USA}
\author{Jon P. Camden}
\email{jon.camden@nd.edu}
\affiliation{Department of Chemistry and Biochemistry, University of Notre Dame, Notre Dame, Indiana 46556, USA}
\affiliation{Department of Materials Science and Engineering, University of Tennessee, Knoxville, Tennessee 37996, USA}
\author{David J. Masiello}
\email{masiello@uw.edu}
\affiliation{Department of Chemistry, University of Washington, Seattle, Washington 98195, USA}

\begin{abstract}
In this Letter, we exploit recent breakthroughs in monochromated aberration-corrected scanning transmission electron microscopy (STEM) to resolve infrared plasmonic Fano antiresonances in individual nanofabricated disk-rod dimers. Using a combination of electron energy-loss spectroscopy (EELS) and theoretical modeling, we investigate and characterize a subspace of the weak coupling regime between quasi-discrete and quasi-continuum localized surface plasmon resonances where infrared plasmonic Fano antiresonances appear. This work illustrates the capability of STEM instrumentation to experimentally observe nanoscale plasmonic responses that were previously the domain only of higher resolution infrared spectroscopies.
\end{abstract}
\maketitle

Since the pioneering work of Ruthemann in 1941 \cite{ruthemann1941diskrete}, inelastic electron scattering experiments using collimated electron beams have made enormous advances in their ability to simultaneously combine and correlate spectroscopic information with spatial imaging at the nanoscale. Today, electron energy-loss spectroscopy (EELS) performed in a monochromated aberration-corrected scanning transmission electron microscope (MAC STEM) can resolve energy losses below 5 meV, with a focused fast electron probe that possesses qualities similar to an ultrafast, near-field, white light source and is only a few atoms in diameter. Paired with modern developments in instrumentation, these properties of the electron probe have made possible the simultaneous spectroscopy and nanometer-scale imaging of optically bright and dark electronic, and even vibrational excitations in nanoparticles \cite{abajo10,Masiello12a,Krivanek2014,lagos2017mapping,lourencco2017vibrational,C3CS60478K,wu2017probing,doi:10.1146/annurev-physchem-040214-121612,koh2011high,duan2012nanoplasmonics,scholl2013observation}, plasmonic energy and charge transfer \cite{p15,doi:10.1021/acs.jpclett.6b01878,doi:10.1021/acs.nanolett.5b00802}, and magneto-optical metamaterials \cite{Fan28052010,Ogut2012toroidal,cherqui14,NL16,doi:10.1021/acsphotonics.8b00519}, heralding a new frontier of materials discovery that is inaccessible to far-field optical spectroscopies.

Despite these advances, the asymmetric Fano lineshape \cite{PhysRev.124.1866}, first observed in 1959 in the EEL autoionization spectrum of He gas \cite{lassettre1959collision,lassettre1964inelastic}, remains elusive in the EELS of plasmonic systems. In his seminal 1961 work \cite{PhysRev.124.1866}, Fano interpreted the observed lineshapes in terms of a configuration interaction between Helium's discrete $2s2p$ double electronic excitation and the scattering continuum. In recent years, so-called Fano interferences or antiresonances have been observed in a variety of optical \cite{RevModPhys.82.2257,nano17,nanolett17,natphot16,rybin2015switching,limonov2017fano,lu2012plasmonic}, plasmonic \cite{simoncelli2018imaging,doi:10.1021/nl2021366,wang2018tunable,lovera2013mechanisms,shafiei2013subwavelength,lopez2012fano,verellen2014mode,hao2008symmetry}, and transport \cite{gores2000fano,PhysRevLett.105.056801,rotter2004tunable} experiments that involve weak coupling between spectrally narrow and broad resonances as generalizations of Fano's original discrete and continuum states. Theory has debated the ability of EELS to capture the Fano antiresonance in plasmonic systems \cite{bigelow13a,PhysRevB.90.155419,doi:10.1021/acsphotonics.5b00416}, providing impetus for a careful experimental investigation.

Motivated by a new generation of STEM monochromators, we construct and measure the spectral response of a plasmonic nanostructure that satisfies two critical requirements for the Fano antiresonance: (1) the individual plasmonic ``configurations'' are weakly coupled to each other, and (2) there is roughly a factor of ten or greater between the linewidths of each configuration, corresponding to the discrete and continuum channels of Fano's original analysis.  These requirements are achieved through the design of a gold disk-rod dimer possessing a series of sharp, experimentally resolvable mid-infrared Fano antiresonances arising from the perturbative influence of the rod's spectrally narrow infrared Fabry-P\'erot (FP) surface plasmon polariton (SPP) resonances \cite{rossouw2011multipolar,nicoletti2011surface,rossouw2013plasmonic,doi:10.1021/nl801504v,martin2014high,Masiello19unp} upon the comparably broad dipole plasmon of the disk. We also present an analytical model that generalizes the Fano lineshape to account for the finite linewidth of both broad (quasi-continuum) and narrow (quasi-discrete) modes, as well as the inherently lossy nature of the interaction between rod and disk modes through the electromagnetic field. Finally, we apply the model to the experimentally measured dimer spectra, showing that it explains the observed features in terms of the incoherent interaction between the rod and disk plasmons in rationally-designed dimers of variable disk diameter and rod length.

\begin{figure}
\includegraphics{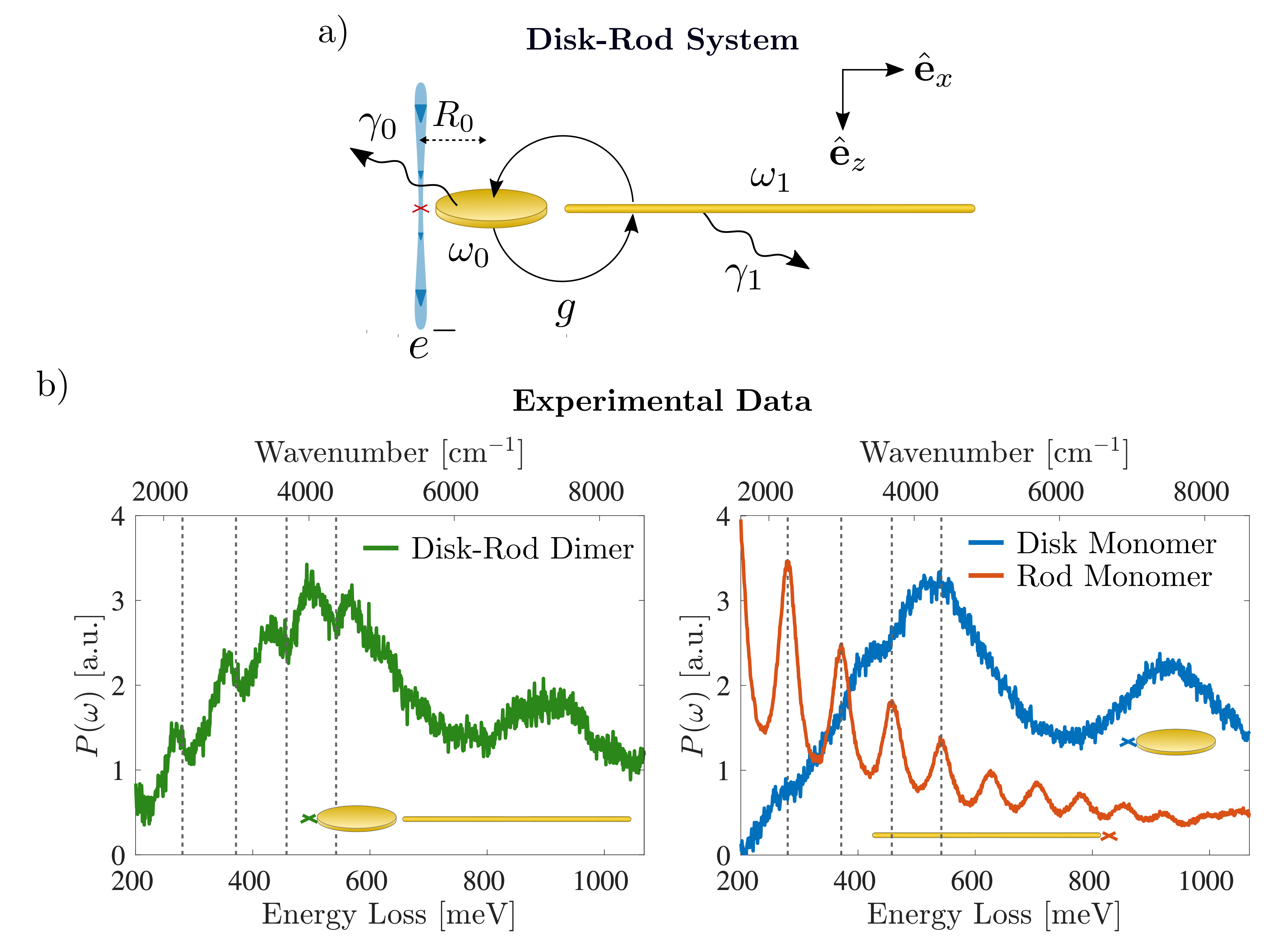}
\caption{\label{f1} (a) Schematic of a gold disk-rod dimer indicating the relevant system parameters and electron-beam location where spectra are acquired (red $\times$). (b) Experimental EEL spectrum of a dimer consisting of a 800 nm diameter gold disk and a 5 $\mu$m long gold rod separated by a 50 nm gap (green curve). Blue and red curves show the monomer spectra for a near-identical disk and rod, respectively. The dimer spectrum is not a simple sum of the two monomer spectra, but instead exhibits a narrow dip at the spectral location of each rod mode (dashed lines). A typical example of the EEL spectrum acquired at the rod end may be found in the Supplemental Material \cite{SI_lollipop}.}
\end{figure}
Fig. \ref{f1}a shows a schematic of the coupled disk-rod system studied, designed such that the dipole plasmon resonance of the disk spans a progression of narrow FP rod modes of alternating parity. Tuning the rod length controls the number of rod modes that overlap with the disk dipole, while both the rod length and disk diameter together determine the degree of spectral overlap between disk and rod modes. Weak coupling is achieved at relatively large disk-rod separations ($\sim50$ nm edge-to-edge), with the parameters necessary for Fano antiresonances falling into a subset of this space where, in addition, there is a factor of $\sim10$ or greater between the disk dipole plasmon and FP rod resonance linewidths. Extensive preliminary experimental and theoretical studies were performed to optimize the plasmon energies and linewidths of the disk and rod monomers such that the disk-rod dimers meet these criteria while retaining the smallest detuning possible between the disk dipole and lower-order rod modes.

The top panel of Fig. \ref{f1}b shows the point EEL spectrum of a disk-rod dimer composed of an 800 nm diameter disk and a 5 $\mu$m rod separated by a 50 nm gap, measured at a beam location 10 nm radially outward from the disk edge along the axis of the rod (green $\times$). For comparison, the bottom panel of Fig. \ref{f1}b displays the EEL spectra for an isolated disk (blue curve) and rod (red curve) of the same size, collected at beam locations indicated by the blue and red $\times$, respectively. The disk monomer spectrum (Fig. \ref{f1}b) reveals a broad resonance around 500 meV attributed to the dipolar disk mode, while the rod monomer spectrum shows a succession of spectrally narrow FP SPP resonances beginning around 200 meV. As anticipated, the spectrum of the coupled system collected on the disk end is not a simple sum of the two monomer spectra, but instead follows the Lorentzian-like ``envelope'' of the isolated disk dipole peak with narrow asymmetric dips at the spectral location of each rod mode, indicative of weak coupling.

Analysis and interpretation of measured EEL spectra is facilitated by analytical modeling of the disk-rod dimer. Considering only the interaction between a single FP mode of the rod with the dipole plasmon of the disk, the surface plasmon resonance solutions of Maxwell's equations can be mapped onto the following set of coupled harmonic oscillators \cite{cherqui14,doi:10.1146/annurev-physchem-040214-121612},
\begin{equation}
\begin{split}
\label{eom}
&\ddot{p}_0+\gamma_{\textrm{nr}}\dot{p}_0- \frac{2e^2}{3m_0c^3}\dddot{p}_0 + \omega_0^2{p}_0-\sqrt{\frac{m_1}{m_0}}\int^t_{-\infty}dt' g(t-t'){p}_1(t')=\frac{e^2}{m_0}E_{\textrm{el}}^x({\bf 0},t)\\
&\ddot{p}_1+\gamma_{\textrm{nr}}\dot{p}_1 + \gamma_{\textrm{rad}}\dot{p}_1 +\omega_1^2p_1-\sqrt{\frac{m_0}{m_1}}\int^t_{-\infty}dt' g(t-t'){p}_0(t')= 0.
\end{split}
\end{equation}
Here $p_i$ labels the $x$-oriented surface plasmons of the disk ($i=0$) and rod ($i=1$) of natural frequency $\omega_i$, nonradiative dissipation rate $\gamma_{\textrm{nr}}$, and effective mass $m_i$ \cite{cherqui14,doi:10.1146/annurev-physchem-040214-121612}. Radiation-reaction forces have been included to account for radiative losses by the system, which in the frequency domain can be repackaged into the total dissipation rates $\gamma_0(\omega) = \gamma_{\textrm{nr}} + 2e^2\omega^2/3m_0c^3$ for the disk dipole mode \cite{jackson} and $\gamma_1 = \gamma_{\textrm{nr}} + \gamma_{\textrm{rad}}$ for the rod mode; here $\gamma_{\textrm{rad}}$ has been used in place of the frequency-dependent Larmor rate due to the non-dipolar nature of the rod modes, which are sufficiently spectrally narrow such that $\gamma_{\textrm{rad}}$ is well-approximated as frequency-independent.

The disk dipole plasmon is driven by the electric field ${\bf E}_{\textrm{el}}({\bf x},t)=-e({\bf x}-{\bf R}_0-{\bf v}t)/\gamma_L^2[(z-vt)^2+(R/\gamma_L)^2]^{3/2}$ of the fast electron moving uniformly with velocity ${\bf v}=\hat{\bf e}_zv$ evaluated at the center of the disk, taken to be the origin. Here $\gamma_L=[1-(v/c)^2]^{-1/2}$ is the Lorentz contraction factor, $\mathbf{R}_0 = -\hat{\bf e}_xR_0$ the electron beam position (Fig. \ref{f1}a red $\times$), and $R=\sqrt{(x+R_0)^2+y^2}$ is the lateral distance between electron probe and field observation point in the impact plane ($z=0$). Due to the relatively large disks studied ($\gtrsim650$ nm in diameter), the rod modes are not directly driven by the evanescent field of the electron when the electron probe is positioned at the disk end of the dimer. No EEL signal is observable above the background when the disk is removed, illustrating the disk's role as an antenna that transfers energy from the electron probe to the rod. 

The coupling strength between the disk and rod plasmon modes depends upon the relative separation and orientation of the disk and rod as well as their respective polarizabilities. In the frequency domain, the coupling is characterized by the complex parameter $g(\omega)$, arising from the interaction energy $U_{\textrm{int}}=-{\bf E}_1\cdot{\bf p}_0$, where ${\bf E}_1$ is the induced electric field of the rod mode evaluated at the disk dipole center. The real part of $g(\omega)$ defines the rate of energy transfer between the disk and rod plasmon modes, while the imaginary part accounts for the lossy nature of this interaction and is related to the degree of interference between the fields of the coupled modes \cite{SI_lollipop}. Because the rod modes are spectrally narrow, the real part of the coupling strength $g(\omega)$ may be treated as approximately frequency-independent. Likewise, the imaginary part is taken to be linear in $\omega$ as $g(\omega)$ is purely real for static fields (i.e., $\omega=0$) and therefore does not have a frequency-independent contribution. Lastly, only the coupled plasmon dynamics oriented parallel to the rod's long axis need be considered due to the high aspect ratio of the rod, justifying the use of the quasi-one dimensional dynamical equations in Eq. (\ref{eom}) with all other collective electronic motion occurring at much higher energy.

The EEL probability $P(\omega)$ per unit frequency $\omega$ of transferred quanta between electron beam and target is obtained by computing the work done on the electron probe by the field induced in the polarized target \cite{ritchie57},
\begin{equation}
\label{Pexpr}
P(\omega)=\frac{|\tilde{E}_{\textrm{el}}^x({\bf 0},\omega)|^2}{\pi\hbar}\textrm{Im}\left[\frac{e^2}{m_0}\left(\omega_0^2 - \omega^2 -i\omega\gamma_0 - \frac{g^2}{ \omega_1^2 - \omega^2 - i\omega\gamma_1}\right)^{-1}\right],
\end{equation}
while the EEL probability for the isolated disk $P_0(\omega)$ is obtained from the above expression by taking $g=0$. The ratio between $P(\omega)$ and $P_0(\omega)$ at the same beam position ${\bf R}_0$ can be cast into the reduced form
\begin{equation}
\label{reduced}
\frac{P(\omega)}{P_0(\omega)}=\left(1 + \textrm{Im}\left[\frac{g^2/\omega\gamma_0}{\omega_1^2 - \omega^2 - i\omega\gamma_1}\right]\right) \Big|\frac{q+\epsilon}{\epsilon+i}\Big|^2
\end{equation}
which generalizes Fano's original lineshape to account for dissipation in both broad and narrow plasmon resonances as well as complex coupling. Here $q(\omega)=(\Omega^2(\omega)-\omega_1^2 +i\omega\gamma_1(\omega))/\omega\Gamma(\omega)$ and $\epsilon(\omega)=(\omega^2-\Omega^2(\omega))/\omega\Gamma(\omega)$ are respectively the complex-valued asymmetry function and reduced frequency expressed in terms of the modified frequency $\Omega^2(\omega)=\omega_1^2- \textrm{Re}[g^2(\omega_0^2 - \omega^2 - i\omega\gamma_0)^{-1}]$ and linewidth $\Gamma(\omega)=\gamma_1(\omega)+ (1/\omega)\textrm{Im}[g^2(\omega_0^2 - \omega^2 - i\omega\gamma_0)^{-1}]$ of the spectral feature described by the interaction of disk dipole and rod plasmon modes. For true Fano antiresonances, the function $q(\omega)\approx q(\omega_1)$ is approximately constant and represents the asymmetry parameter originally proposed by Fano to distill the physics of the antiresonance into a single number that depends upon the basic system parameters \cite{PhysRev.124.1866}. Here, since both disk and rod modes are dissipative, the asymmetry parameter generalizes to a complex-valued number, the real part of which characterizes the degree of asymmetry of the antiresonance. It is important to note that without the second term proportional to $\gamma_1$, $q(\omega)$ would be real-valued and the reduced EEL probability spectrum in Eq. (\ref{reduced}) would vanish at those frequencies where $\epsilon(\omega)=-q(\omega)$ \cite{nanolett17}. However, this is not observed experimentally at any coupling strength due to the finite linewidth of the spectrally narrow rod resonances. Lastly, the standard form of the Fano lineshape is scaled by a frequency-dependent prefactor which accounts for the additional non-disk dissipation channels of the dimer.

Since each rod has multiple plasmon modes that spectrally overlap the disk dipole plasmon resonance, the EEL probability is further generalized as

\begin{equation}
\label{Pexprmulti}
\begin{split}
P(\omega)&=\frac{|\tilde{E}_{\textrm{el}}^x({\bf 0},\omega)|^2}{\pi\hbar}\textrm{Im}\Big[\Big(\omega_0^2 - \omega^2 - i\omega\gamma_0-\sum_j \frac{g_j^2}{\omega_j^2 - \omega^2 - i\omega\gamma_j}\Big)^{-1}\Big]\\
\end{split}
\end{equation}
and the reduced EEL probability may be cast into the approximate form,
\begin{equation}
\label{product}
\begin{split}
\frac{P(\omega)}{P_0(\omega)}&\approx {\cal F}_1\big(q_{1}(\omega),\epsilon_1(\omega)\big){\cal F}_2\big(q_{2}(\omega),\epsilon_2(\omega)\big)\cdots {\cal F}_N\big(q_N(\omega),\epsilon_N(\omega)\big),
\end{split}
\end{equation}
where ${\cal F}_j(q_j(\omega),\epsilon_j(\omega))$ is the Fano lineshape describing the interaction between the $j$th rod plasmon mode and the disk dipole plasmon mode (labeled by the subscript $0$) given by Eq. (\ref{reduced}). This product factorization of the reduced spectrum, which allows for an estimate of the asymmetry function $q_{j}(\omega)$ for each individual rod plasmon mode, is approximate as the rod resonances overlap weakly, causing their individual contribution to the dimer spectrum to depend upon neighboring rod modes through their mutual interaction with the disk dipole plasmon. Nonetheless, the exact form of the reduced EEL spectrum inferred from Eq. (\ref{Pexprmulti}) can be used to demonstrate the accuracy of the simple product form in the weak coupling regime when all rod modes are well-separated spectrally \cite{SI_lollipop}. Lastly, while the model parameters (including $g_j$) could be obtained by approximating the disk and rod by oblate and prolate spheroids and adding the contributions from radiation damping, doing so adds little additional insight into the measurements; thus we obtain these parameters by numerically fitting the experimental spectra.

Measured EEL spectra are collected at ${\bf R}_0$ for a set of fabricated gold disk-rod dimers of varying rod length and disk diameter. All system parameters ($\omega_0$, $m_0$, $\omega_j$, $\gamma_j$, and $g_j$) are obtained for each dimer by least-squares fitting the analytic form for $P(\omega)$ defined by Eq. (\ref{Pexprmulti}) to the spectra. The nonradiative (Drude) dissipation rate of the disk dipole is set prior to fitting according to the value for gold at optical frequencies ($\hbar\gamma_{\textrm{Au}}=69$ meV \cite{sonnichsen2001plasmons}). Initial guesses for the natural frequency $\omega_0$ and effective mass $m_0$ of the disk plasmon are estimated for each dimer by fitting the measured EEL spectra collected at ${\bf R}_0$ of an isolated disk, while initial guesses for $\omega_j$ and $\gamma_j$ of the $N$ rod plasmons are estimated from the EEL spectra of an isolated rod. As a check of the fitting procedure, the parameters obtained from each dimer spectrum are used to reconstruct the disk monomer spectrum $P_0(\omega)$, rod monomer spectrum $P_{\textrm{rod}}(\omega) = \sum_j P_j(\omega)$ (where $P_j(\omega)$ is identical in form to $P_0(\omega)$ with indices interchanged where appropriate), and the reduced EEL probability spectrum $P(\omega)/P_0(\omega)$ for each structure. We note that, while any spectrum can be fit by an arbitrary collection of oscillators, the approach here is restricted by the number of oscillators present in the monomer spectra.

\begin{figure}
\includegraphics{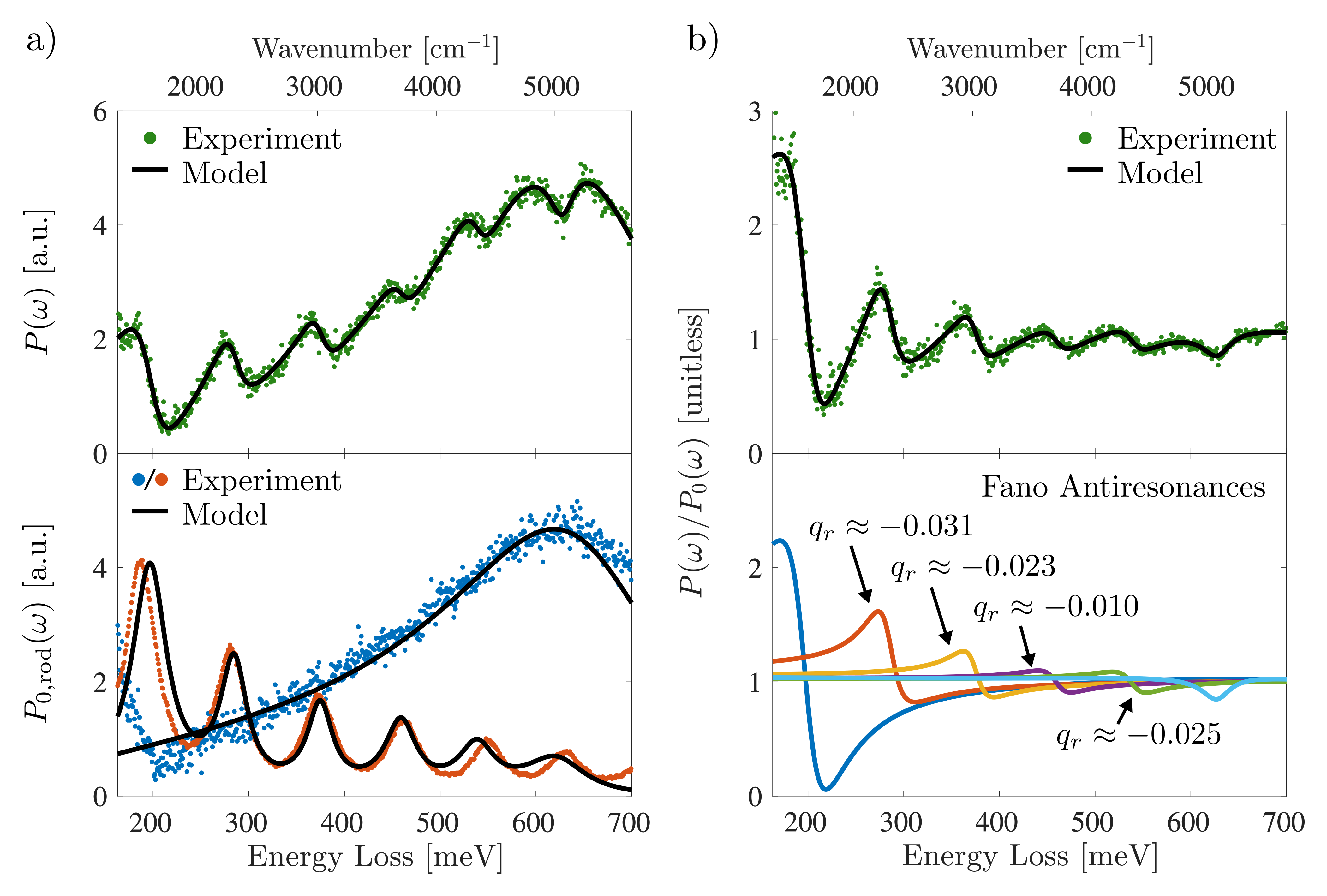}
\caption{\label{f3}EEL point spectrum of a gold disk-rod dimer composed of a 650 nm diameter disk and a 5 $\mu$m rod separated by a 50 nm gap. The spectrum exhibits a progression of infrared Fano antiresonances due to the interaction between the broad disk dipole plasmon resonance and the spectrally narrow plasmon modes of the rod. The upper panels display the (a) experimental (green) and fit (black) EEL spectrum and (b) reduced EEL spectrum of the dimer collected at the disk end. The lower panel of (a) shows the experimental monomer spectra of an isolated disk (blue) and rod (red). As an independent check of the fitting procedure, the theoretical monomer spectra are reconstructed from the dimer fit parameters (black curves), showing excellent agreement. The lower panel of (b) displays the decomposition of each antiresonance in the reduced spectrum into a product of Fano lineshapes ${\cal F}_j(q_{j},\epsilon_j)$ as described in Eq. (\ref{product}) with the corresponding value of the real part of the asymmetry parameter $q_{r,j}=\textrm{Re}\,q_j(\omega_j)$ indicated above each feature.}
\end{figure}
Fig. \ref{f3} shows the result of this analysis for a dimer composed of a 650 nm diameter gold disk and a 5 $\mu$m gold rod separated by a 50 nm gap. The dimer point EEL spectrum, collected at a beam position 10 nm radially outward from the disk edge (green $\times$), is shown in the upper panel of Fig. \ref{f3}a (green bullets) with the fit to Eq. (\ref{Pexprmulti}) overlaid (black curve). The bottom panel of Fig. \ref{f3}a compares the experimental EEL spectra obtained from a 650 nm disk monomer (blue bullets) and a 5 $\mu$m rod monomer (red bullets) to the theoretical monomer spectra reconstructed from parameters obtained from fitting the dimer spectrum (black curve). Due to small geometrical variations between the isolated monomer rods and disks versus those which compose the dimers, the monomer spectra will not, in general, exactly match those corresponding to the dimer disk and rod. In addition, deviation between the reconstructed and experimental disk monomer spectra is expected on the higher-energy side of the disk dipole peak where the quadrupole plays a non-negligible dynamical role. Despite these limitations, Fig. \ref{f3}a shows excellent agreement between reconstructed and experimental spectra, which further validates our ability to extract the monomer parameters from the dimer spectra. To compare with our theoretical analysis, Fig. \ref{f3}b displays the reduced EEL probability (green bullets) obtained by dividing the experimental spectrum by the theoretically reconstructed isolated disk spectrum $P_0(\omega)$ (top), along with the decomposition into a progression of individual Fano lineshapes ${\cal F}_j(q_{j},\epsilon_j)$ (bottom).

\begin{figure}
\includegraphics{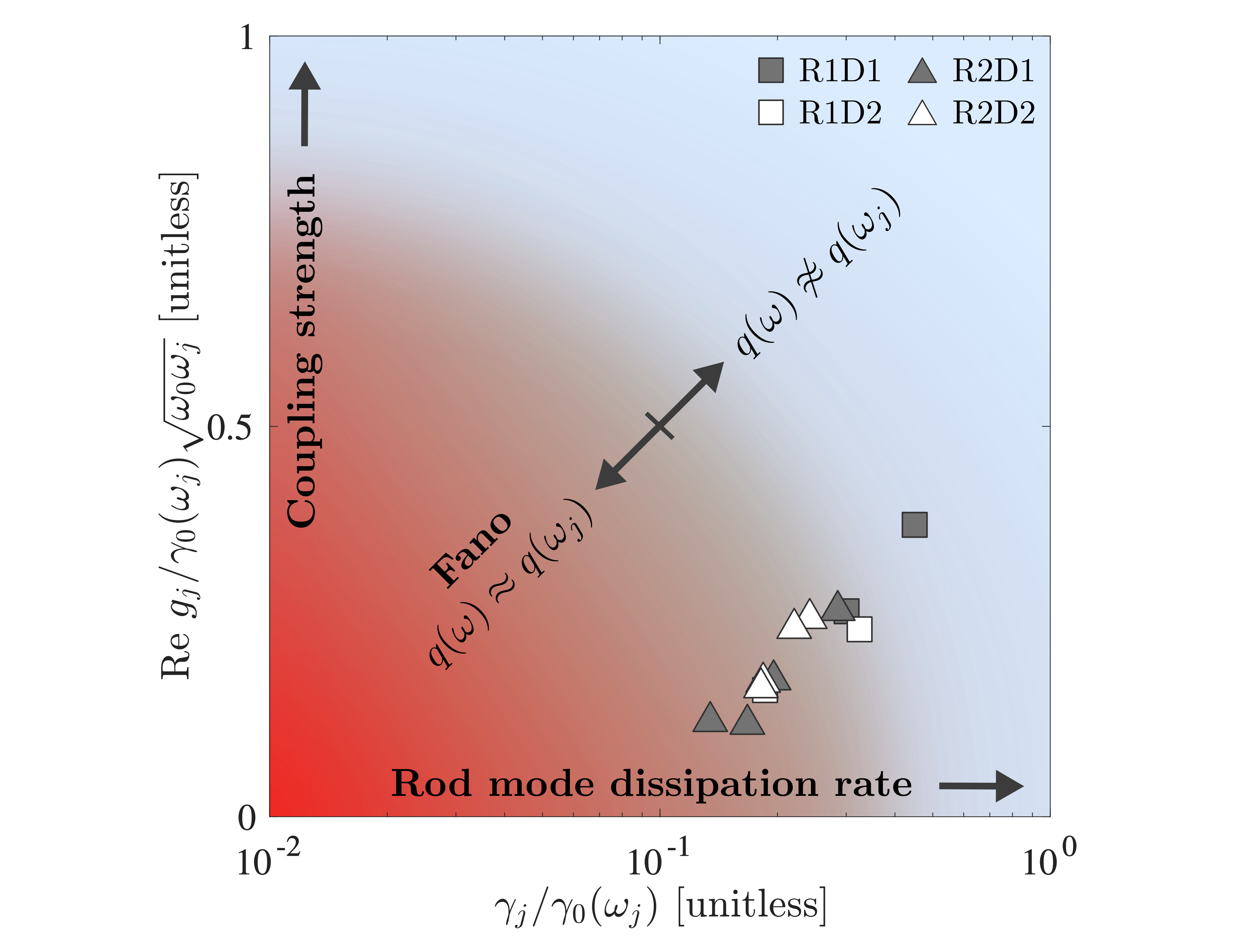}
\caption{\label{f4}Graphical summary of the interaction between individual rod resonances and the disk dipole plasmon in a collection of disk-rod dimers. Each mode pair is represented by a distinct symbol and is characterized by its relative coupling strength $\textrm{Re}\, g_j/\gamma_0(\omega_j)\sqrt{\omega_0\omega_j}$ and dissipation rate $\gamma_j/\gamma_0(\omega_j)$. Dimers denoted by R1 (R2) consist of 2.5 $\mu$m (5 $\mu$m) long rods, while those denoted by D1 (D2) consist of $650$ ($800$ nm) diameter disks. In all dimers, the disk and rod are separated by a gap of 50 nm and since $\textrm{Re}\, g_j/\gamma_0(\omega_j)\sqrt{\omega_0\omega_j}=1$ denotes the boundary between weak and strong coupling, all dimers are in the weak coupling regime. The gray triangle symbols indicate specific Fano antiresonances shown explicitly in Fig. \ref{f3}.}
\end{figure}
This analysis is repeated for a set of four unique disk-rod combinations \cite{SI_lollipop} and summarized in Fig. \ref{f4} to illustrate the variation in coupling strength and relative linewidth as a function of disk and rod size. Underlying each data point is a particular rod FP mode (labeled $j$) which interacts with the disk dipole plasmon (labeled 0). Dimers denoted by D1 (D2) consist of a $650$ ($800$) nm disk while those denoted by R1 (R2) contain $2.5$ ($5$) $\mu$m long rods. In all cases, the disk and rod are separated by a gap of $50$ nm and EEL spectra are collected 10 nm radially outward along the rod long axis from the disk edge (Fig. \ref{f1}a, red $\times$). As each dimer contains multiple overlapping disk and rod modes, these four structures generate 12 modes available for analysis. For all dimers, the lowest and highest energy rod resonances are not included as explicit data points due to uncertainties imposed by subtraction of the zero-loss peak and interactions with the SiO$_2$ substrate phonon mode at lower energies ($\lesssim 200$ meV) and the influence of the disk quadrupole at higher energies ($\gtrsim 650$ meV). The full spectra, however, are displayed in the Supplemental Material \cite{SI_lollipop}.

All disk-rod mode pairs are found to be in the weak coupling regime as each data point satisfies the inequality $\textrm{Re}\, g_j/\gamma_0(\omega_j)\sqrt{\omega_0\omega_j}<1$ \cite{rodriguez2016classical,novotny2010strong}. Additionally, multiple disk-rod mode pairs are found to obey the linewidth condition $\gamma_j\sim\gamma_0/10$ (Fig. \ref{f4} red region), including those highlighted in Fig. \ref{f3}, thus satisfying both requirements for the emergence of Fano antiresonances in the coupled spectrum. Additionally, these results indicate that the size of both the rod and disk play a crucial role in determining whether the disparity in linewidths between the disk and rod modes is sufficient to observe a sharp antiresonance. We find that the longer 5 $\mu$m rods (R2) in combination with the 650 nm diameter disk (D1) optimally balance the two criteria for sharp Fano antiresonances, while supporting a progression of rod modes which are minimally detuned from the disk dipole such that disk-rod interaction is non-negligible.

In conclusion, we resolve for the first time Fano antiresonances in the EEL spectrum of a plasmonic nanostructure. This is achieved by rationally designing a gold disk-rod dimer supporting rod resonances that are spectrally narrow relative to the disk dipole. Observation of the asymmetric lineshapes is facilitated by a new generation of monochromated and aberration-corrected STEMs which open the infrared spectral region to interrogation. We develop a theoretical model which generalizes the original Fano lineshape to account for dissipation in both the quasi-discrete and the quasi-continuum channels in STEM-EELS. This analysis makes explicit the classification of the observed dimer lineshapes in terms of the asymmetry parameter $q,$ as discovered in the autoionization spectrum of He by Fano in 1961 \cite{PhysRev.124.1866}. This combined experimental and theoretical work not only resolves an ongoing discussion in the literature about the existence of Fano lineshapes in the EELS of plasmonic systems \cite{bigelow13a,PhysRevB.90.155419,doi:10.1021/acsphotonics.5b00416}, but also showcases the ability of the latest generation of monochromated STEMs to observe spectrally narrow plasmonic responses that were previously the domain only of higher resolution optical spectroscopies.

\begin{acknowledgments}
We thank Yueying Wu, Xiang-Tian Kong, Zhongwei Hu, and Jacob A. Busche for preliminary experimental and numerical efforts which led to the work presented. This work was supported by the University of Washington, the University of Notre Dame, the University of Tennessee, the U.S. Department of Energy Basic Energy Sciences under Award Number DE-SC0018040 for the theoretical modeling and numerical simulation of the electron probe (K.C.S., S.C.Q., D.J.M.), DE-SC0018169 for the EELS measurements and analysis (A.O., X.H., J.P.C.), and DMR-1709275 (R.C., P.D.R.) for the synthesis of the dimer heterostructures. The STEM experiments and dimer synthesis were conducted at the Center for Nanophase Materials Sciences, which is a DOE Office of Science User Facility (J.C.I.). This research was conducted, in part, using instrumentation within ORNL's Materials Characterization Core provided by UT-Batelle, LLC under Contract No. DE-AC05-00OR22725 with the U.S. Department of Energy, and sponsored by the Laboratory Directed Research and Development Program of Oak Ridge National Laboratory, managed by UT-Battelle, LLC, for the U.S. Department of Energy. 
\end{acknowledgments}

\bibliography{/Users/david/TeX/bib/masiello}

\end{document}